\documentclass[twoside]{sfm}

\input{sf.def}

\begin{document}

\def\llm{{\sc LLmodels}}
\def\atl{{\sc ATLAS9}}
\def\aatl{{\sc ATLAS12}}
\def\starsp{{\sc STARSP}}
\def\aur{$\Theta$~Aur}
\def\logg{\log g}
\def\tauros{\tau_{\rm Ross}}
\def\kms{km\,s$^{-1}$}
\def\bz{$\langle B_{\rm z} \rangle$}
\def\degr{^\circ}
\def\aaps{A\&AS}
\def\aap{A\&A}
\def\apjs{ApJS}
\def\apj{ApJ}
\def\Rh{\rule{20.0pt}{0.0pt}}
\def\Rv{\rule[-0.1in]{0.0pt}{20.0pt}}
\def\Sum{N_{\rm tot}}
\def\rmxaa{Rev. Mexicana Astron. Astrofis.}
\def\mnras{MNRAS}
\def\actaa{Acta Astron.}
\newcommand{\Tef}{T$_{\rm eff}$~}
\newcommand{\Vt}{$V_t$}
\newcommand{\CC}{$^{12}$C/$^{13}$C~}
\newcommand{\CDC}{$^{12}$C/$^{13}$C~}

\pagebreak

\thispagestyle{titlehead}

\setcounter{section}{0}
\setcounter{figure}{0}
\setcounter{table}{0}

\markboth{Cowley \& Hubrig}{The Herbig Ae PDS2}
\renewcommand{\thefootnote}{\fnsymbol{footnote}}
\titl{The Herbig Ae Star PDS2\footnote[5]{Based on Data
obtained from the ESO Science Archive Facility}}
{Cowley, C.$^1$, Hubrig, S.$^2$}
{$^1$University of Michigan, Ann Arbor, MI, USA, email: {\tt cowley@umich.edu} \\
 $^2$Leibnitz-Institute f\"{u}r. Astrophysik, Potsdam Germany}
\renewcommand{\thefootnote}{\arabic{footnote}}
\abstre{We present a preliminary abundance analysis
of the isolated Herbig Ae star PDS2 (CD -53 251,
2MASS J01174349-5233307).  Our adopted model has \Tef
= 6500K, $\log(g) = 3.5$. It is likely that PDS2
belongs to the class of young stars with abundances
resembling the $\lambda$ Bootis stars. }

\baselineskip 12pt

\section{Introduction}
PDS2 was noted in the Pico dos Dias (PDS) survey of
the IRAS Point Source Catalog of objects as a
possible Herbig Ae or Be star.  Vieira, et al.
\cite{vieira} discuss the PDS survey in addition to
giving photometric data as a part of their large
study of HAe/Be candidate stars.  They assign a
distance of 340 pc, and give the colors U$-$B = 0.01,
and B$-$V = 0.38.  Sartori, et al. \cite{sartori}
included PDS2 in their analysis of Herbig Ae/Be
candidates, and classified its SED as Group 2, with
``far-IR emission corresponding to intermediate
values of IR excess.''  Bernabi, et al.
\cite{bernabei}, and Marconi, et al. \cite{marconi},
have studied pulsations of PDS2, while Hubrig, et al.
\cite{hubrig09} announced a detectable magnetic
field.

In addition to these interesting properties, PDS2 is
a young object at quite a high galactic latitude with
no obvious relation to a star-forming region.
Nevertheless, its claim to youth is based on its
infrared colors and mass accretion rate (Pogodin, et
al. \cite{pogodin}).

\section{Observations of the present study}

\begin{figure}[!t]
\begin{center}
\hbox{
 \includegraphics[width=8cm,angle=270]{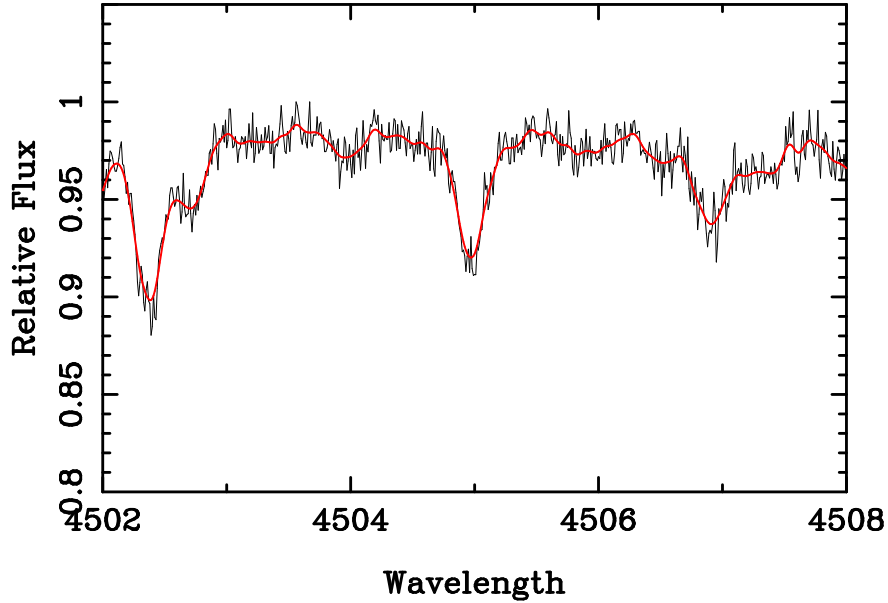}
}
\vspace{-5mm}
\caption[]{{\rule{0.0pt}{20.0pt}}Averaged (black) and averaged+filtered spectra (gray/red).}
\label{fig:plfint}
\end{center}
\end{figure}

This work is based primarily on HARPS spectra of PDS2 obtained from the ESO archive.  We downloaded two sets of nine spectra, obtained on the nights of 11/12  November 2008.  An additional nine were obtained on 13 November 2008.
On each night, the spectra were taken within a few hours of one another.  We attempted to allow for the small differences in radial velocity, but ultimately concluded that straight averages (for each night) were better.  The averaged spectra were still rather noisy.  The
spectra from 11 December were clearly of higher quality
than those from 13 November;  all measurements were
therefore based on the former set of spectra.  We worked from
a Fourier-filtered spectrum, illustrated in Fig.~\ref{fig:plfint}.

In addition to the HARPS spectra, we used X-shooter spectra obtained on 28 June 2010 by
Pogodin, et al. \cite{pogodin}.

\section{Fixing the atmospheric parameters}\label{lab:params}

As a part of their work on accretion, Pogodin, et al. \cite{pogodin} made an effort to fix the model atmospheric parameters for PDS2.  They concluded that \Tef = 7000K, $\log(g) = 4.0$, and $v \sin(i) = 30$ \kms.  It is particularly straightforward to refine the atmospheric parameters of stars in this temperature-gravity range as:

\begin{itemize}
\item the Balmer lines are very sensitive to \Tef, and
nearly insensitive to gravity;
\item the lines of species such as Fe I, Cr I, or Ni I
are similarly insensitive to gravity;
\item one can determine the Fe abundance from weak Fe I (or other neutral metallic) lines, and the microturbulence, $\xi_t$, from strong 
Fe I lines; and
\item the surface gravity then follows from corresponding lines of the second spectra (e.g. Fe II, Ni II, etc.).
\end{itemize}

\begin{figure}[!t]
\begin{center}
\hbox{
 \includegraphics[width=8cm,angle=270]{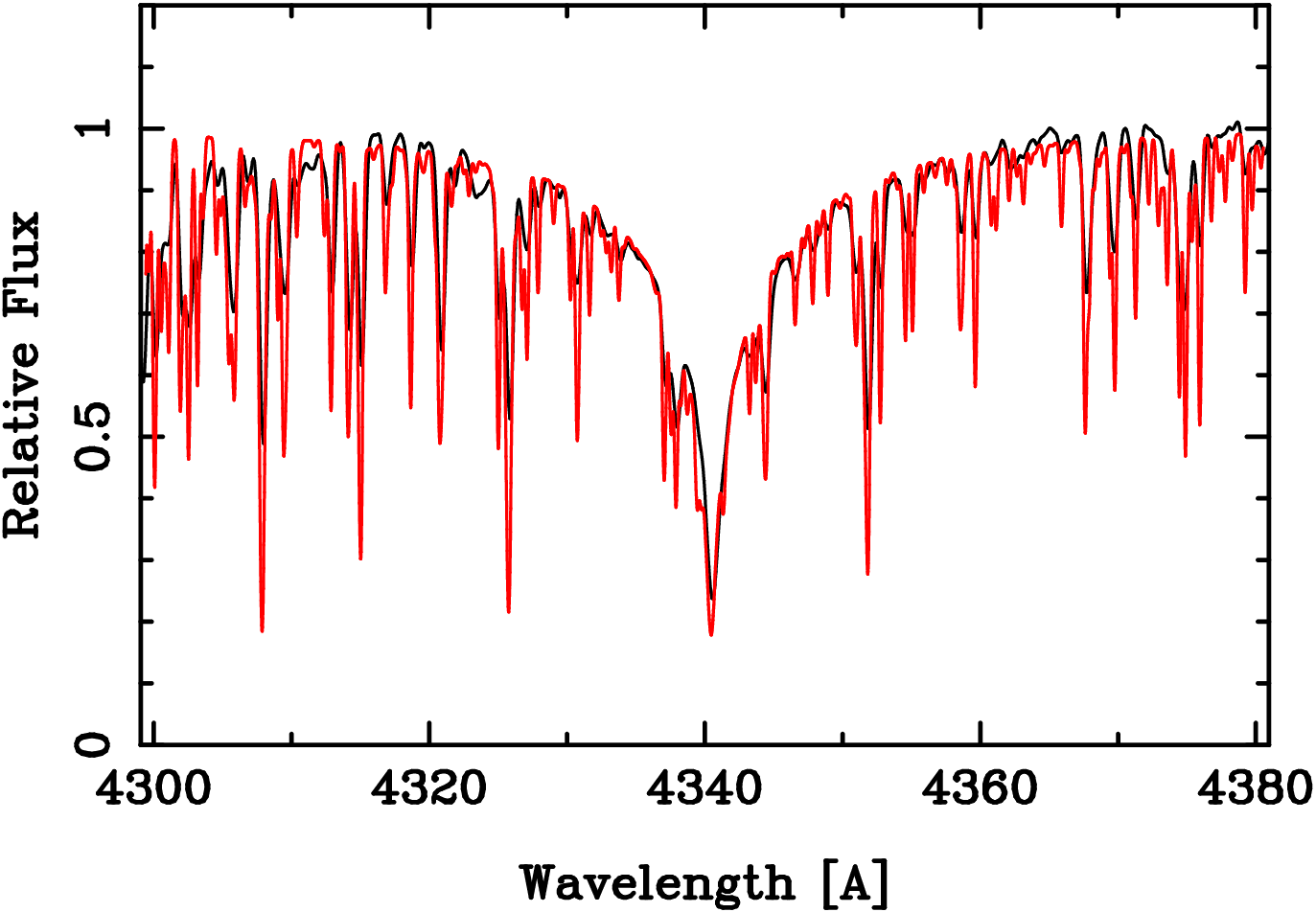}
}
\vspace{-5mm}
\caption[]{{\rule{0.0pt}{20.0pt}}Observed (x-shooter,
(black) and calculated (red/gray) profile for \Tef = 6500K.}
\label{fig:gammax}
\end{center}
\end{figure}

Very good agreement for the Balmer profiles was obtained from models with \Tef = 6500 K, both from HARPS and x-shooter spectra.  The problem of normalization of echelle spectra was palliated somewhat by the narrower profiles of the relatively cool PDS2, compared to hotter Herbig Ae stars.  Additionally, the part of the profiles most sensitive to the temperature is the near wing, where
the line depths vary from ca. 0.15 to 0.4.  A calculated
H$\gamma$ profile for \Tef = 6600K is slightly too strong, but
perhaps acceptable.  The fit for \Tef = 6700 is unacceptable.

An overall check on the atmospheric parameters is to 
see that abundances from weak lines of the first and 
second
spectra of well-determined species agree (see Tab.~\ref{tab:abs})
From the iron and titanium equilibria,
we conclude that PDS has a somewhat lower gravity than
Procyon--we adopt $\log(g) = 3.5$.

\begin{table}
\begin{center}
\caption{\label{tab:aabbs}Abundances of several elements
in PDS2.\label{tab:abs}}
\begin{tabular}{l  c  c  r c c} \hline
El{\Rv}  & $\log(N/\Sum)$ &$\pm$(sd)&n&$(\log(N/{\Sum})_\odot$   &[$N$] \\
\hline
C I &$    -3.53$&0.14& 3&$ -3.61 $&$    +0.08  $\\   
S I &$    -4.85$&0.06& 6&$ -4.92 $&$    +0.07  $ \\  
Ti I&$    -7.08$&0.16& 6&$ -7.09 $&$    +0.01   $\\  
Ti II&$   -7.21$&0.07& 3&$ -7.09 $&$    -0.11  $ \\  
Cr I&$    -6.52$&0.31&12&$ -6.40 $&$    -0.12  $\\   
Cr II&$   -6.49$&0.06&11&$ -6.40 $&$    -0.09  $\\   
Fe I&$    -4.83$&0.09& 8&$ -4.54 $&$    -0.29  $ \\ 
Fe II&$   -4.76$&0.13&17&$ -4.54 $&$    -0.22  $ \\
Zn I&$    -7.76$&0.13& 3&$ -7.48 $&$    -0.28  $\\
\hline
\end{tabular}
\end{center}
\end{table}

\section{The metal lines}\label{sec:metal}

\begin{figure}[!t]
\begin{center}
\hbox{
 \includegraphics[width=8cm,angle=270]{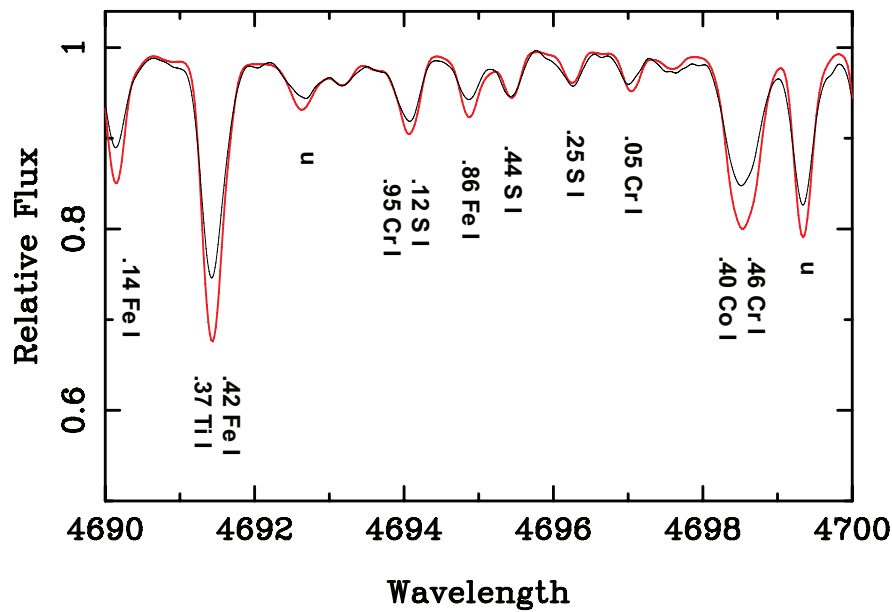}
}
\vspace{-5mm}
\caption[]{{\rule{0.0pt}{20.0pt}}Metal and neutral sulfur 
lines in Procyon (red/gray) and PDS2 (black).  Decimal
fractions of wavelengths (in Angstroms) are given.
}
\label{fig:sulfu2}
\end{center}
\end{figure}

Fig. ~\ref{fig:sulfu2} contrasts a HARPS spectrum
PDS2 (black) with a UVESPOP archiveal spectrum
\cite{bagnulo} of Procyon (red/gray).  The Procyon
spectrum was convoluted with a $v \sin(i) = 10$ \kms
rotational profile to approximately match the
observed PDS spectrum.

Generally, the metallic lines in Procyon are
stronger, indicating slightly larger abundances, as
is found from an equivalent width analysis.  Because
the numerous neutral lines are insensitive to
gravity, and the temperatures of the two stars are
nearly the same, within the errors, this is a direct
indication of relative metal weakness in PDS2. Note
that the two S I lines not influenced by blending are
{\it not} weaker in PDS2. This again comports with
the quantitative analysis, and the overall assessment
that PDS2 shows mild $\lambda$ Boo characteristics,
since sulfur is a volatile element.

\section{Conclusions}

The abundances in PDS2 depart mildly from solar.
However, they do so systematically, showing mild
depletions of refractory elements.  The volatile
elements have normal abundances, with the exception
of zinc, which is clearly depleted.  This zinc
anomaly is a problem for theories that suggest
volatility as a key to the abundance pattern. The
same difficulty holds for the $\lambda$ Boo stars, as
noted, for example, by Heiter \cite{heiter}.

\bigskip
{\it Acknowledgements.} This research has made use of the
SIMBAD and VALD databases, as well as the NIST online
Atomic Spectroscopy Data Center.  CRC thanks his 
colleagues at Michigan for many useful conversations
concerning young stars.



\begin{thebibliography}{99}
\bibitem{bagnulo}
{Bagnulo, S, Jehin, E., Ledoux, C., et al.} 2003, The
Messenger, 114, 10.

\bibitem{bernabei}
{Bernabei, S., Marconi, M., Ripepi., V., et al.}
2007, Commun. asteroseismol., 150, 57.

\bibitem{heiter}
{Heiter, U.} 2002, A\&A, 381, 959.

\bibitem{hubrig09}
{Hubrig, S., Stelzer, B., Sch\"{o}ller, M., et al.}
2009, A\&A, 502, 283.

\bibitem{marconi}
{Marconi, M., Ripepi, V., Bernabei, S, et al.} 2010,
Astrophys. Space Sci., 328, 109.

\bibitem{pogodin}
{Pogodin, M. A., Hubrig, S., Yudin, R. V., et al.} 2012,
Astron. Nachr., 333, 594.

\bibitem{sartori}
{Sartori, M. J., Gregorio-Hetem, J., Rodrigues, C. V.
et al.} 2010, AJ, 139, 27.

\bibitem{vieira}
{Vieira, S. L. A., Corradi, W. J., B., Alencar, S. H. P, et al.} 2003, AJ, 126, 2971.



\end{thebibliography}
\end{document}